\documentclass[twocolumn,showpacs,preprintnumbers,amsmath,amssymb]{revtex4}
\usepackage{graphicx}
\usepackage{dcolumn}
\usepackage{bm}
\usepackage{color}
\begin{document}


\title{A percolation model with continuously varying exponents}
\author{R. F. S. Andrade$^{1}$, H. J. Herrmann$^{2,3}$}
\affiliation{$^1$Instituto de F\'{i}sica,
Universidade Federal da Bahia,
40210-210, Salvador, Brazil.\\
$^{2}$Computational Physics, IfB, ETH-H\"{o}nggerberg, Schafmattstr. 6,
8093, Z\"{u}rich, Switzerland.
\\$^{3}$Departamento de F\'{i}sica, Universidade Federal do Cear\'{a},
Campus do Pici, 60455-760, Fortaleza, Brazil.}

\date{\today}

\begin{abstract}
{This work analyzes a percolation model on the diamond hierarchical lattice (DHL), where the percolation transition is retarded by the inclusion of a probability of erasing specific connected structures. It has been inspired by the recent interest on the existence of other universality classes of percolation models. The exact scale invariance and renormalization properties of DHL leads to recurrence maps, from which analytical expressions for the critical exponents and precise numerical results in the limit of very large lattices can be derived. The critical exponents $\nu$ and $\beta$ of the investigated model vary continuously as the erasing probability changes. An adequate choice of the erasing probability leads to the result $\nu=\infty$, like in some phase transitions involving vortex formation. The percolation transition is continuous, with $\beta>0$, but $\beta$ can be as small as desired. The modified percolation model turns out to be equivalent to the $Q\rightarrow1$ limit of a Potts model with specific long range interactions on the same lattice. }
\end{abstract}

\maketitle

\section{Introduction}

In the past years there have been interesting discussions about the possible existence of percolation phenomena \cite{Stauffer1994} with unusual phase transitions \cite{Achlioptas2009,Andrade2009,Araujo2010a,Chen2011,Araujo2011}. While the usual bond percolation model is based on purely random occupation of still empty bonds, several new models have been proposed with some kind of restriction for the placement of a new bond. A common feature of the quoted works (and many others in the recent literature) is to add rules that favor the inclusion of bonds between sites that do not increase the largest cluster size, and reduce the occupation probability of spanning links, i.e., links that if occupied would cause spanning \cite{Schrenk2012a,Cho2013}. This leads to two main consequences: i) the percolation transition is retarded towards a larger critical value of ($p=p_c$), where $p$ is the probability occupation of an individual site; ii) a much sharper transition is observed when the new value $p_c$ is reached, since any connecting bond sharply increases the largest cluster.

Such investigations are heavily based on numerical simulations, which makes it difficult to uncover the actual nature of the new transition. Also, as the new simulation rules require a global knowledge of the system, it becomes complicated to translate them into a model that can be treated analytically. Nevertheless, it is now accepted that the original \emph{explosive percolation} model follows a second order transition from the non-percolating to the percolating phase \cite{Nagler2010, Costa2010, Riordan2011}. Let us remind that the existence of sharp transitions has been recently investigated in the context of attacks (bond or node removal) in coupled complex networks stressing the importance of understanding the role of some key elements in this broad class of systems \cite{Stanley2010}.

The present work should be regarded in the context of the above discussion. In first place, it investigates the effects related to retarding and modifying the nature of the percolation transition caused by the inclusion of new rules in a well known percolation model. Next, it makes a contribution to understanding the role of a few key elements in the origin of sharp transitions. The adopted approach is amenable to analytical treatment, since it is based on changes in the rules of a standard bond percolation model on hierarchical lattices. Due to their geometrical scale invariance, exact analytical methods based on renormalization methods can be derived. Although we restrict our results to the diamond hierarchical lattice (DHL) \cite{Migdal1975a,Migdal1975b,Kadanoff1976,Berker1979,Bleher1979,Kauffman1981}, this strategy can be extended to more complex geometries.

For the bond percolation model on DHL \cite{Stanley1977,Stanley1978,Nakanishi1981,Tsallis1996}, the probability $p_0$ of a bond being occupied is kept constant while the lattice grows. In turn, it is possible to derive exact maps for the probability that the root sites are connected in subsequent generations, say $p_{g+1}=p_{g+1}(p_g)$. After iterating the map, the result $p_{\infty}(p_0)=1$ ($p_{\infty}(p_0)=0$) indicates if the two root sites of the lattice are connected (disconnected).

The DHL results are approximations to an alternative percolation process on the square lattice. Here, we would start with a four site square and a given bond probability occupation, and construct the lattice by putting together four equivalent units at each step. However, since this procedure is not exact for Euclidian lattices, the random bond occupation process is the preferred method. The hierarchical assembly method does not allow to randomly choose a subset of bonds and decide which of them should be added as in the quoted models. If one wants to describe recent advances in percolation studies on hierarchical structures, it becomes necessary to devise an alternative procedure to avoid the emergence of large clusters.

Our proposal is to add an erasing probability to the usual percolation model when we go from one generation to the next. Therefore, the new system is characterized by an overall \emph{reduction} in the number of bonds as compared to the results of the original system (i.e., without any erasing probability) for the same value of $p_0$. The same is also valid for the average number of sites in the percolating cluster. This contrasts with models based on the choice of a new bond among a preselected random subset. Here, the value of $p$ after inserting any given number of bonds is exactly the same as in the corresponding original percolation model \cite{Achlioptas2009,Cho2013}.

We also discuss that, for a particular choice of erasing probability, the results for the percolation model are reproduced by a $Q$-state Potts model \cite{Potts1952} (in the usual $Q\rightarrow 1$ limit) provided we include an extra set of competing bonds between root sites of each generation. Such results are derived within a transfer matrix (TM) approach, which has been used to investigate uniform and non uniform spin models on hierarchical structures \cite{Andrade1993,Andrade1999,Andrade2005,Araujo2010b}.

The rest of this paper is so organized: in Section 2 we define the model, and derive the recurrence maps for the percolation probability between the root sites and for the number of connected sites in the percolating cluster. Section 3 discusses analytical and numerical results, emphasizing the derivation of the critical exponents and the influence of the erasing probability. In Section 4 we derive a Potts model with long range interactions that is equivalent, in the $Q\rightarrow 1$ limit, to the explored percolation model. A transfer matrix (TM) method is used to obtain thermodynamical and magnetic properties, which coincide with those derived from the percolation model. Finally, Section 5 closes the paper with concluding remarks.

\section{The percolation model}

Any geometrically hierarchical structure can be constructed in a sequence of steps (or generations $g$), by replacing a given geometrical element in the $g$-th generation by a more complex structure in the $g+1$-th generation. In the case of the DHL, we start at $g=0$ with a line segment linking two root sites ($r_1,r_2$). For any $g \geq 1$, we replace each bond of the previous generation $g-1$ by a set of $r$ parallel branches, with $s-1$ inner sites in each one of them. In this work, we consider $r=s=2$ (see Fig.1). The resulting self-similar graph has a fractal dimension $d_f=\log rs / \log s$ \cite{Tsallis1996}. The maximal number of bonds, sites, and shortest distance between root sites depend on $g$. They will be denominated, respectively, as $\mathbf{B}_g=4^g$, $\mathbf{N}_g=2(4^g+2)/3$, and $\mathbf{D}_g=2^g$.

\begin{figure}
\includegraphics*[angle=-90,width=8cm]{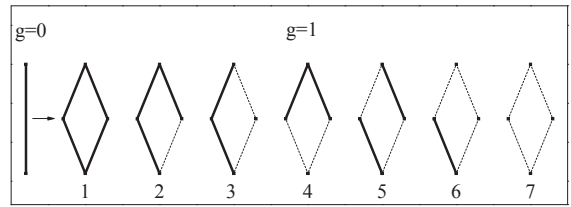}
\caption{First step of construction of the usual percolation model on the DHL: solid and dotted lines indicate occupied and non-occupied bonds. At $g=1$, seven possible degenerated different configurations exist, which can be percolating (1-3) or non-percolating (4-7).}\label{fig0}
\end{figure}

It is well known that the results for spin models on the DHL are equivalent to approximations produced for Euclidian lattices by the Migdal-Kadanoff real space renormalization group (RG) \cite{Kadanoff1976,Berker1979,Kauffman1981,Tsallis1996}. Adopting the same point of view, the percolation model we investigate here can also be regarded as an approximation to a similar percolation problem on the square lattice. The usual percolation model starts by assigning, at $g=0$, the probability $p_0$ that the root sites are connected. Let $p_{g}(p_0)$ denote the probability that, at generation $g$, the two root sites are connected. It is straightforward to derive the following recurrence maps expressing $p_{g+1}$ and $q_{g+1}$ in terms of $p_g$ and $q_g=1-p_{g}$, $g=0,1,...$ :

\vspace{-0.2cm}

\begin{equation}\label{eq0}
 p_{g+1} = p_g^4 + 4p_g^3q_g + 2p_g^2q_g^2
\end{equation}

\vspace{-0.5cm}

\begin{equation}\label{eq1}
 q_{g+1} = 4p_g^2q_g^2 + 4p_gq_g^3 +  q_g^4 .
\end{equation}

\noindent  Each term in the above equations expresses the contribution to $p_{g+1}$ or $q_{g+1}$ of a given configuration formed by different percolating and non-percolating structures at generation $g$ (see Fig. \ref{fig0}). For instance, the term $p_g^4$ indicates the contribution of the configuration 1, formed by four percolating structures in the generation $g$.

The probability $p_{g+1}$ can be reduced if we multiply any of the three terms on the r.h.s. of Eq. (\ref{eq0}) by constant factors (say $A,B,$ and $C$), with $0\leq A,B,C \leq 1$ and $A+B+C<3$. This corresponds to terms $\sim A-1, B-1, C-1 \leq 0$ that are added to Eq. (\ref{eq0}), while corresponding terms $\sim 1-A, 1-B, 1-C \geq 0$ are added to the r.h.s. of Eq. (\ref{eq1}), increasing the value of $q_g$. With these modifications, the two maps become

\vspace{-0.2cm}

\begin{equation}\label{eq1b}
 p_{g+1} =  A p_g^4 + 4 B p_g^3q_g + 2 C p_g^2q_g^2
\end{equation}

\vspace{-0.5cm}

\begin{equation}
\begin{array}{c}\label{eq1c}
 q_{g+1} = 4p_g^2q_g^2 + 4p_gq_g^3 +  q_g^4 + \\\\
 (1-A) p_g^4 + 4 (1-B) p_g^3q_g + 2 (1-C) p_g^2q_g^2.
\end{array}
\end{equation}

The parameters $A, B,$ and $C$ impact differently on the maps. If we consider the individual effect of each parameter, condition $A<1,$ $B=C=1$ causes most severe changes in the behavior of the model, followed by $B<1,$ $A=C=1$. The reason is that $A<1$ reduces the probability of having highly populated configurations with four percolating structures in the previous generation $\sim p_g^4$. To simplify our analysis, we consider from now on that only one of the three parameters is taken to be less than 1. Therefore, unless explicitly indicated, $A<1$ also requires $B=C=1$, with similar assumptions being valid when we state $B<1$ or $C<1$.

To illustrate which processes are described by the erasing action at $g=1$ (see Fig. \ref{fig0}), a value $C<1$ amounts to replace the percolating configuration 3 of Fig. \ref{fig0}, formed by 2 bonds, by other non-percolating configurations (4-7), which may have 2, 1 or 0 bonds. If it is replaced by a 2-bond structure (configurations 4 or 5 of Fig. \ref{fig0}), the action is close to what is done in the original model by Achlioptas et. al \cite{Achlioptas2009}. However, for the percolation transition, it is not relevant to know the actual configuration of the new non-percolating structure. Indeed, at $g=2$ the values of $p_2$ and $q_2$ depend only on $p_1$ and $q_1$, not on specific configurations. Because of this, when $A<1$ or $B<1$, it is not crucial to indicate which non-percolating structure at generation $g$ replaces the percolating one in the evaluation of $p_{g+1}$.

Two other measures are relevant for a more precise characterization of the percolation process: the average number of bonds and the mass of the largest cluster. In this work we will consider the latter measure, which is understood as the number of sites connected to the root sites. Therefore, we define $I_{p,g}$ (and $\overline{I}_{q,g}$) as the normalized average \emph{internal} mass of the largest connected cluster in a percolating (non-percolating) configuration. We emphasize that, to have a simpler form of the recurrence maps, at any generation $g$ the \emph{internal} mass does not include the two root sites. Recurrence maps can be derived to describe the dependence of these functions for the $g+1$-th generation in terms of the corresponding values at generation $g$.  After identifying the contributions of the proper configurations, it is possible to derive the recursion relations for the average mass of the largest connected cluster in the percolating and non-percolating regions as

\vspace{-0.5cm}

\begin{equation}\label{eq4}
\begin{array}{c}
 I_{p,g+1} = \frac{1}{p_{g+1}}  [(2k_g+4\ell_gI_{p,g})p_g^4 +
(2k_g+\ell_g(3I_{p,g} + \\\\
\overline{I}_{q,g}))4p_g^3q_g +
(k_g+\ell_g(2I_{p,g}+\overline{I}_{q,g}))2p_g^2q_g^2  ]
\end{array}
\end{equation}

\vspace{-0.2cm}

\begin{equation}\label{eq4a}
\begin{array}{c}
 \overline{I}_{q,g+1} = \frac{1}{q_{g+1}}
[(2k_g+\ell_g(2I_{p,g}+2\overline{I}_{q,g}))4p_g^2q_g^2 + \\\\
(2k_g+\ell_g(I_{p,g}+2\overline{I}_{q,g}))4p_gq_g^3 +
2\ell_g\overline{I}_{q,g}q_g^4 ],
\end{array}
\end{equation}

\noindent where $k_g=1/\mathbf{N}_{g+1}$ and $\ell_g=\mathbf{N}_{g}/\mathbf{N}_{g+1}$. The recurrence maps for $I_{p,g}$ and $\overline{I}_{q,g}$ do not depend explicitly on $A,B,$ and $C$. However, the resulting iterated values are influenced by these parameters through $p_g$ and $q_g$. To obtain the normalized average mass of the percolating and non-percolating clusters at generation $g$ with the inclusion of the root sites we consider, respectively, $\mathcal{M}_{p,g}=2/\mathbf{N}_g + I_{p,g}$ and $\overline{\mathcal{M}}_{q,g}=2/\mathbf{N}_g + \overline{I}_{q,g}$. Finally, the average mass of sites connected to the root sites is expressed by $M_g = p_g \mathcal{M}_{p,g} + q_g\overline{\mathcal{M}}_{q,g}$. \noindent As we will show in the next Section, the maps (\ref{eq1b}) and (\ref{eq4}) lead to transition properties that depend on the value of $A$ and $B$.

\section{Results}

We start this Section by revising the critical properties of the usual percolation model on hierarchical lattices \cite{Stanley1977,Stanley1978,Nakanishi1981,Tsallis1996}. By imposing the fixed point (FP) condition $p_{g+1}=p_g=p_c$ on Eq. (\ref{eq0}), it is possible to derive a $4-$th degree polynomial equation $P(p_c)=p_c^4-2p_c^2+p_c=0$ with roots: $p_{c,1}=-(\sqrt{5}+1)/2$, $p_{c,2}=0$, $p_{c,3}=(\sqrt{5}-1)/2$, $p_{c,4}=1$, where $p_{c,1}$ has no physical meaning. $p_{c,2}$ and $p_{c,4}$ correspond to the attractive non-percolating and percolating solutions, while the critical properties are related to the unstable $p_{c,3}$. If the maps (\ref{eq0}) are iterated starting from the points $p_0=p$, the system evolves for the percolating (non-percolating) phase when $p>p_{c,3}$ ($p<p_{c,3}$). By eliminating $q$ in Eq. (\ref{eq0}) and linearizing the resulting equation in the neighborhood of $p=p_{c,3}$, we obtain the eigenvalue $\lambda=6-2\sqrt{5}$. The critical exponent $\nu$, which governs the divergence of the correlation length at the critical point, can be expressed in terms of $\lambda$ by \cite{Stauffer1994,Stanley1977} $\nu=\log s/\log \lambda=1.6352...$, where $s=2$ according to the discussion in Sec. II.

\begin{figure}
\includegraphics*[angle=-90,width=8cm]{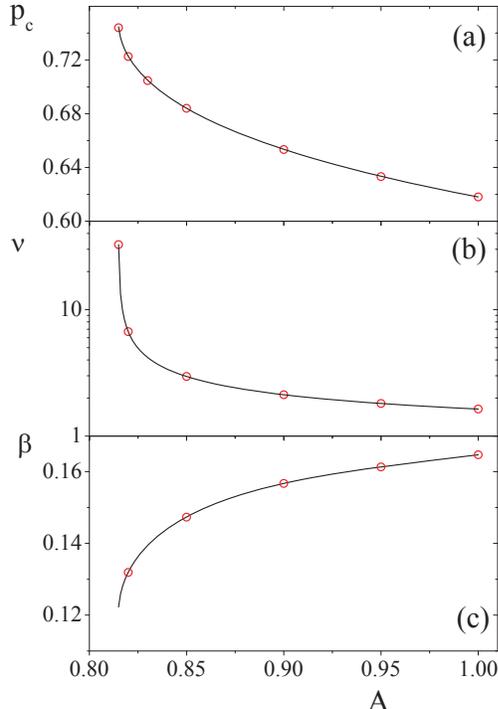}
\vspace{-1.5cm}
\caption{a) Dependence of $p_c=p_{c,3}$ on $A$ (solid line). The red dots were obtained by making the correspondence between the critical temperature $T_c$ of the modified $Q=1$ Potts and the value of $p_c$ (see Sec. IV and Fig. \ref{fig2}). b) Dependence of $\nu$ on $A$. The solid curve follows from the evaluation of $p_{c,3}$ and $\lambda$. The red circles indicate the values obtained from the slopes of $\log \xi(T)$ as function of $\log[(T-T_{c})/T_{c}]$, where $\xi$ is the correlation length of the equivalent $Q\rightarrow1$ Potts model (see Sec. IV and inset in Fig. \ref{fig2}). c) Dependence of $\beta$ with respect to $A$. The solid curve follows from the evaluation of $p_{c,3}$ and $p_{c,4}$. As in (b), the red circles indicate the values obtained from the dependence of the spontaneous magnetization $m(T)$ of the $Q\rightarrow1$ Potts model as function of $T$ (see Sec. IV and Fig. \ref{fig60}). They coincide with values obtained from scaling analysis of $M$ as function of $p-p_{c,3}$ in a neighborhood of $p_{c,3}$. }\label{fig4}
\end{figure}

If we consider the condition $A<1$, it turns out from the structure of the 4-th degree FP equation that $p_{c,2}=0$ is still a solution, but the other three roots of $P(p_c)=0$ can not be given by simple analytical expressions as before. Of course they can expressed with the help of the Cardan's formulae, or can be evaluated by numerical methods. We take a shortcut, and look for the roots $p^*$ of the derivative $dP/dp=3p^2(A-2) +2|_{p=p^*}=0$. If they are real, they represent extreme points that must be necessarily between the roots of $P(p_c)=0$. It easily follows that $p^*=\pm \sqrt{2/(6-3A)}$. Indeed, for $A=1$, the positive root $p^*=\sqrt{2/3}$ lies between $p_{c,3}$ and $p_{c,4}$. In general, $P(p^*_+)=4p^*_+/3$. We find that $p_{c,3}$ and $p_{c,4}$ are real as long as $P(p^*_+)\le 1$.

This provides the following conditions on $A$: if $A>A_s=22/27$, the system admits three physical FP's: $p_{c,2}=0$, which corresponds to the non-percolating state; $p_{c,3}>(\sqrt{5}-1)/2$, the threshold value for the emergence of the percolation phase; and $p_{c,4}<1$, which describes percolating state. When $A$ decreases from $1$ to $A_s$, the roots $p_{c,3}$ and $p_{c,4}$ approach each other and finally coalesce at $p_{c,3}(A_s)=p_{c,4}(A_s)=0.75$. Finally, the non-percolating root $p_{c,2}=0$ is the only attracting set of Eq. (\ref{eq1b}) if $A<A_s$. The dependence of the percolation transition expressed by $p_{c,3}$ as function of $A$ is illustrated in Fig. \ref{fig4}a.

The above described FP properties for $A \in [22/27,1)$ have two direct physical consequences: i) the percolation transition occurs at a larger value $p_{c,3}$; ii) even if $p>p_{c,3}$ there exists a small probability that the percolating cluster fails to emerge, since the solution of the maps (\ref{eq1b}) is attracted to $p_{c,4}<1$. This behavior is different from usual percolation models.

The value of $\lambda$ at $p_{c,3}$ decreases monotonically with $A$, reaching the value $\lambda=1$ at $A=A_s$. The absolute value of the attractive eigenvalue of the linearized map in the neighborhood of $p_{c,4}$ also decreases with $A$. If $A\gtrsim A_s$ and $p\gtrsim p_{c,3}$, the trajectory formed by the values of $p_g$ requires a larger number of generations to depart from the neighborhood of $p_{c,3}$. Such dependence of $\lambda$ on $A$ causes an increase of $\nu$ as $A$ decreases. In fact, it is possible to show that, in a neighborhood of $A_s$, $\nu(A)\sim (A-A_s)^{-1/2}$. Thus, the percolation phase transition has a \emph{singular} behavior at $A=A_s$. The dependence of $\nu$ on $A$ is shown in Fig. \ref{fig4}b. The result $\nu=\infty$ indicates that the correlation length diverges exponentially in a neighborhood of $p_c$. Such behavior has been observed in other systems, like planar rotors, $XY$ and vertex ice models on the square lattice \cite{Berezinskii,Kosterlitz1973}. It is associated with the presence of an essential singularity in the singular part of the free energy, and non-linear evolution equations of the RG approach. The value $\lambda=1$ leads to a different behavior of the iterates of $p$ when $A=A_s$. In this case, if we insert $p_g=p_{c,3}+\delta_g$ into Eq. (\ref{eq1b}), the map reduces to $\delta_{g+1} = p_{g+1}-p_{c,3} = \delta_g - 2\delta_g^2$, so that $\delta_{g+1}$ depends on the second degree term $\delta_g^2$ in any infinitesimal neighborhood of $p_{c,3}$. As a consequence, the trajectory moves towards $p_{c,3}$ if $p>p_{c,3}$, but moves away from it when $p<p_{c,3}$.

\begin{figure}
\includegraphics*[angle=-90,width=8cm]{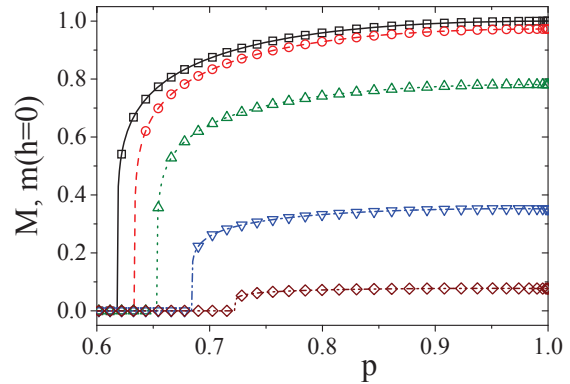}
\caption{Dependence of the size of the largest cluster $M$ and the spontaneous magnetization $m$ of the $Q\rightarrow1$ Potts model that is equivalent to the $A$ model (see Section IV). Lines indicate the behavior of $M_g$ for $A=1$ (black solid), 0.95 (red dashes), 0.90 (green dots), 0.85 (blue dash-dot), and 0.82 (wine dash dot dot). Symbols (squares, circles, up-triangles, down triangles, and diamonds) correspond to values of $m_g$ for the same values of $A$.}\label{fig60}
\end{figure}

The largest cluster mass $M=\lim_{g\rightarrow\infty}M_g$ as function of $p$ for $A_s<A\leq1$ is illustrated in Fig. \ref{fig60}. The curves follow from the iteration of the maps (\ref{eq1b}) and (\ref{eq4}). The value of $M$ depends on whether $p<p_{c,3}$ or $p>p_{c,3}$. In the first case, the only possible FP solution of (\ref{eq4}) is $I_{p}*=\overline{I}_{q}*=0$, so that $M=0$. When $p>p_{c,3}$, $M$ depends not only on the properties of $p_{c,4}$ but also on transient steps, which are the values of $g$ for which $p_g$ is not yet in a sufficiently close neighborhood of $p_{c,4}$. 

When $A=1$, $q_{c,4}=1-p_{c,4}=0$ but, for $A_s<A<1$, it is observed that $p_{c,4}<1$. This fact changes the nature of the possible FP solutions of the maps (\ref{eq4}). Indeed, if both $p_{c,4}$ and $q_{c,4}$ are non-zero, the only exact FP solution is $I_{p}*=\overline{I}_{q}*=0$. Nevertheless, the numerical iteration of the maps (\ref{eq4}) indicates that the convergence to $I_{p}*=\overline{I}_{q}*=0$ is extremely slow. This is illustrated in Figure 3 where we draw several curves for $M_{g=200}$ and different values of $A$. For this value of $g$ the lattice contains already a very large number of sites, $(\mathbf{N}_g \sim 10^{120})$, which largely exceeds the accepted number of baryons in the universe. There we clearly see that $0.1 \lesssim M_g \lesssim 1$ when $p>p_{c,3}$, except in the immediate neighborhood of the transition point. It is also possible to recognize the decrease of $M_g(p)$ as $A$ decreases, which can exemplarily be measured through $M_{g=200}(p=1)$. It is important to notice that, when $A_s<A<1$, $M_{g}$ receives two non-zero, non-equivalent contributions ($p_{c,4} \mathcal{M}_{p,\infty}$ and $q_{c,4}\overline{\mathcal{M}}_{q,\infty}$), while it depends only on $\mathcal{M}_{p,g\rightarrow\infty}$ when $A=1$.

The critical behavior of $M$, which is observed for $p_0>p_{c,3}$, follows from the analysis in the neighborhood of $p_{c,3}$ of Eqs. (\ref{eq4}) and (\ref{eq4a}) as well as Eqs. (\ref{eq1b}) and (\ref{eq1c}). In the first place, it amounts to replace the non-linear maps (\ref{eq4}) and (\ref{eq4a}) by a linear system described by the matrix

\vspace{-0.5cm}
\begin{equation}\label{eq22}
    \Omega=\left(
      \begin{array}{cc}
        p_{c,3}(1+p_{c,3}-p_{c,3}^2) & p_{c,3}(1-p_{c,3}^2)/2 \\
        p_{c,3}(1-p_{c,3}^2) & (1+p_{c,3})(1-p_{c,3}^2)/2 \\
      \end{array}
    \right),
\end{equation}

\noindent with real eigenvalues $\omega_1<1$ and $\omega_2$, such that $|\omega_2|<\omega_1$. While $p_g$ stays in the neighborhood of $p_{c,3}$ ($\delta_g\ll1$), the evolution of (\ref{eq4}) and (\ref{eq4a}) is dominated by $\omega_1$, so that we obtain

\vspace{-0.5cm}
\begin{equation}\label{eq23}
    M_g \sim \omega_1^g\,\, \mathrm{or}\,\, M_g\simeq E \omega_1^g.
\end{equation}

\noindent Since we still consider the restriction $A_s<A<1$, the linearization of Eqs. (\ref{eq1b}) and (\ref{eq1c}) leads to $\delta_g \sim \lambda ^g \delta_0$. Let $g^\dag$ be an integer such that, if $g>g^\dag$, the condition $\delta_g\ll1$ no longer holds and the approximations $\delta_g \sim \lambda ^g \delta_0$ and $M_g \sim \omega_1^g$ do not provide accurate solutions to the maps. If we express $g^\dag$ in terms of $\delta_{g^\dag}=\lambda^{g^\dag}\delta_0$, the largest magnitude of $\delta_g$ where the linear evolution is valid, we obtain $g^\dag(\delta_0) = \log(\delta_{g^\dag}/\delta_0)/\log \lambda$. Of course the choice of $\delta_{g^\dag}$ impacts the precision with which Eq. (\ref{eq23}) is fulfilled but, as we will see, the expression for $\beta$ does not depend on $\delta_{g^\dag}$.

To continue with our analysis, we define $g^{\ddag}\gg g^\dag$ by the condition that, if $g>g^{\ddag}$, $|p_g-p_{c,4}|\ll 1$. It follows that, when $g>g^{\ddag}$, $M_g$ depends on the eigenvalue $o_1$ of a matrix $O$, which describes the linearized evolution of Eqs. (\ref{eq4}) and (\ref{eq4a}) in the neighborhood of $p_{c,4}$. $O$ is obtained after replacing $p_{c,3}$ by $p_{c,4}$ in Eq. (\ref{eq22}). Under this condition, the solution for $M_g$ satisfies $M_{g+G}/M_g = o_1^G$.

Now let us write down the usual scaling behavior for $M$ close to $p_{c,3}$, where Eq. (\ref{eq23}) is valid. If we consider two nearby values $\delta_0$ and $\delta_0'$, it follows that

 \vspace{-0.5cm}
\begin{equation}\label{eq24}
    \frac{M(\delta_0)}{M(\delta_0')}=
    \left(\frac{\delta_0}{\delta_0'}\right)^\beta
    \simeq\omega_1^{[g^\dag(\delta_0)-g^\dag(\delta_0')]},
\end{equation}

\noindent where we have used, as a first approximation, $M(\delta_0) \simeq E \omega_1^{g^\dag}$. Note that we used the largest value of $g$ for which Eq. (\ref{eq23}) still holds. Taking the logarithm on both sides of Eq. (\ref{eq24}), and expressing $g^\dag(\delta_0)$ in terms of the logarithms of $\delta_{g^\dag},\, \delta_0,$ and $\lambda$, we are lead to $\beta=-\log\omega_1/\log \lambda$. This expression is still not completely correct as it does not take into account the influence of $p_{c,4}$ on $\beta$. Such influence is correctly dealt with if we consider the value of $M(\delta_0)$ for which the number of iterations that are performed when $g>g^{\ddag}$ equals $g^\dag$. In other words, we assume that the maps are iterated the same number of steps in the immediate neighborhoods of $p_{c,3}$ and $p_{c,4}$. This amounts to divide $M(\delta_0)$ by $o_1^{g^\dag}$, so that $M(\delta_0) \simeq E (\omega_1/o_1)^{g^\dag}$. With this accurate treatment, we are lead to the correct estimate $\beta=-\log(\omega_1/o_1)/\log \lambda$, which is valid for $A\leq 1$.

The dependence of the critical exponent $\beta$ as function of $A$ is illustrated in Fig. \ref{fig4}c. There we draw the values of $\beta$ after the evaluation of $\omega_1$, $o_1$, and $\lambda$. We superpose also a few values of $\beta$ evaluated by taking the numerical derivatives of $\log M_g$ as function of $\log p$ in the neighborhood of $p_{c,3}$. The perfect agreement between the two evaluation methods corroborates the scaling arguments developed above, allowing for the rapid evaluation of $\beta$ by for all $A_s<A\leq 1$. We remark that the direct evaluation of $\beta$ becomes very difficult for $A \sim A_s$, as a very large number of iteration becomes necessary for $p$ to leave the immediate neighborhood of $p_{c,3}$ and approach $p_{c,4}$. Concomitantly, the value of $M$ becomes smaller and smaller due to the fact that $o_1<1$, making the direct search for the scaling behavior of $M$ very elusive. Finally, at $A \equiv A_s$, we have $p_{c,3}=p_{c,4}=3/4$, what allows to obtain the exact value $\beta=(-8+926\sqrt{7753})/21 \cong 0.119838...$.

The non-zero values of $\beta$ for all intervals of interest indicate that the order parameter of the percolation transition always increases continuously from $M=0$ at $p_{c,3}$. The universality class changes, but the second order character remains the same.

We carried out a similar analysis for the $B\leq1$ model. The results have some similarities to those for $A\leq1$. When $B$ decreases, the percolation transition occurs at a larger value of $p_{c,3}$, which moves in the positive direction towards $p_{c,4}=1$. This behavior prevails until a critical value $B=B_s=3/4$ such that, if $B<B_s$ the percolation phase vanishes. Then $p_{2,c}=0$ becomes the only stable FP with a clear physical meaning. As in the $A\leq1$ model at $A=A_s$, the critical exponent $\nu$ diverges at $B_s$. It can be shown that, for both conditions, $\xi\sim \exp[|p_c-p|^{-1}]$. The major difference between the two conditions ($A<1$ and $B<1$) is the fact that $p_{c,4}$ does not decrease with $B$, but stays fixed at its original value. $p_{c,3}$ reaches $p_{c,4}=1$ and collapses with it at $B_s$. Contrary to the $A\leq1$ case, however, the exponent $\beta\rightarrow 0$. In fact, if we let $B=3/4 + \delta_B$, it is possible to show that, to leading order in $\delta_B$, $p_{c,3}=1-4\delta_B$, $\nu=\log 2/\log (1+4\delta_B)$, and $\beta=16 \delta_B^2 $. In spite of very small values of $\beta$ when $\delta_B \rightarrow 0$, it does not characterize a discontinuous transition \cite{Achlioptas2009}. Strictly speaking, this occurs only for $B=3/4$ when $\beta = 0$. However, in this case, a finite value of $M$ is observed in the single point $p=1$.

Finally, the behavior of the $C\leq1$ model does not present any qualitative changes with respect to that of usual percolation on the DHL. There is no restriction on the existence of the the critical point $p_{c,3}$ even if the value of $C$ is set to zero.

\section{The equivalent long range interaction Potts model}

Let us now show that it is possible to derive a Potts model which, in the $Q\rightarrow1$ limit, becomes equivalent to the $A\leq1$ percolation model. The general Hamiltonian for a nearest neighbor Potts spin model \cite{Potts1952,Tsallis1996} can be written as

\vspace{-0.5cm}
\begin{equation} \label{eq5}
\mathcal{H}=-\sum\limits_{(i,j)}J_{ij}\delta(\sigma_i,\sigma_j)\
-h\sum\limits_i\delta(\sigma_i\,1) ,
\end{equation}

\noindent where $\sigma _i= 1, 2, ..., Q$ indicates $Q$-state Potts spin variables, $\delta(i,j)$ denotes the Kronecker $\delta$ function, the double sum is performed over pairs of nearest neighbor sites $(i,j)$, and the external field is assumed to point along the $Q=1$ direction. The general form of Hamiltonian (\ref{eq5}) holds for any lattice, including DHL.

\begin{figure}
\includegraphics*[angle=-90,width=8cm]{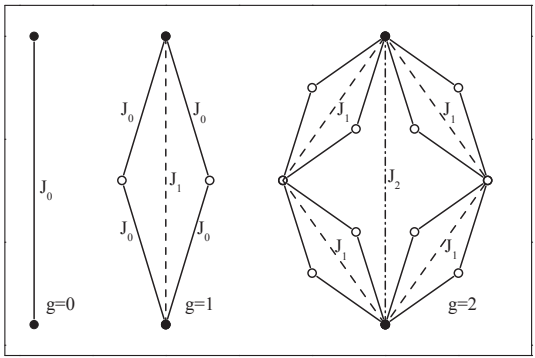}
\caption{Schematic representation of the modified Potts model on the DHL. The dashed and dot-dashed lines for $g=1$ and $g=2$ indicate the extra AF interactions $J_1$ and $J_2$, which are added to the usual nearest neighbor Potts model to account for the erasing of percolating clusters described by $A$. }\label{fig1}
\end{figure}

It is well known that, for the uniform nearest neighbor model with coupling constant $J_{ij}=J_0$, a formal equivalence exists between the bond percolation problem and the Potts model in the $Q\rightarrow1$ limit. Then, the thermodynamical properties of the spin model are equivalent to the results obtained from Eqs. (\ref{eq0}) and (\ref{eq4}), provided the following identification is made

\vspace{-0.5cm}
\begin{equation}\label{eq7a}
 p_0=1-\exp(-J_0/T).
\end{equation}

Consider now modifying the Hamiltonian $\mathcal{H}$ in such a way that it becomes equivalent to the new percolation model in the DHL described by Eq. (\ref{eq1b}), including the observed changes in the nature of the transition between the ordered ($\Leftrightarrow$ percolating) and disordered ($\Leftrightarrow$ non-percolating) states. The search for a suitable modification was based on the fact that, in the modified model, a similar condition as given by Eq. (\ref{eq7a}) should hold. In particular, we note that retarding the percolation transition is equivalent to a decrease in the value of the critical temperature $T_c$. The desired equivalence requires to reduce the ferromagnetic coupling interaction between the spins. This should be done by reducing in a non trivial way the effect of $J_0$. We identified a possible way to obtain this effect, which consists in adding extra anti-ferromagnetic (AF) bonds $J_g$ between the root sites at each generation $g\geq1$, as illustrated in Fig. \ref{fig1}. In this process, the system at generation $g$ consists of four subsystems at generation $g-1$ plus one extra bond coupling the two $g$ root sites. Note that the four $g-1$ subsystems carry along all previously introduced AF bonds, so that the $g$-th system contains exactly $4^{g-g'}$ bonds $J_{g'}$, $g'=0,1,...,g$ (see Fig.1).

Thus we formally define the new Hamiltonian as

\vspace{-0.5cm}
\begin{equation} \label{eq8}
\mathcal{H}=-\sum\limits_{(i,j)}J_0\delta(\sigma_i,\sigma_j)
-\sum\limits_{g=1}^{\infty}\sum\limits_{(i,j)_g}J_g\delta(\sigma_i,\sigma_j)\
-h\sum\limits_i\delta(\sigma_i\,1) ,
\end{equation}

\noindent where $(i,j)_g$ identifies the pairs of spins that are first neighbors when all links introduced at previous generations $0,1,...,g-1$ are erased from the DHL. The new coupling constants $J_g$ will be used to obtain the desired equivalence between percolation and Potts models, which requires that they depend on the erasing probability $A$.

The exact scale invariance of DHL permits the use of a TM formalism for the evaluation of the thermodynamical properties. A detailed description of all steps for the implementation of this method on hierarchical lattices has been discussed in a number of previous works (see, e.g., \cite{Andrade1993,Andrade1999,Araujo2010b}), so that we will indicate only the main necessary steps for its implementation. For the short range $Q$-state Potts model, it amounts to write down $\mathcal{T}_g$, a $Q\times Q$ TM connecting the root sites, which depends only on the $Q^2$ distinct configurations this pair of spins may assume. At zero magnetic field, the matrix $\mathcal{T}_0$ has only two different matrix elements $\forall Q$: $a_0=\exp(J_0/T)$ (where we have set $k_B=1$), and $b_0=1$. For $g>0$, each $\mathcal{T}_g$ element is a partial trace accounting for the Boltzmann weight contributions from all configurations involving the intermediate spins. As long as $J_g=0 \,\, \forall g\geq 1$, the matrix elements of $\mathcal{T}_{g+1}$ can be expressed in terms of those of $\mathcal{T}_g$ by the following nonlinear maps

\vspace{-0.5cm}
\begin{equation}
\label{eq5a}
a_{g+1} = (a_{g}^2+(Q-1)b_{g}^2)^2,
\end{equation}

\vspace{-0.5cm}
\begin{equation}
\label{eq5b}
b_{g+1} = b_{g}^2(2a_{g}+(Q-2)b_{g})^2.
\end{equation}

 \noindent The numerical iteration of Eqs. (\ref{eq5a}-\ref{eq5b}) leads to the partition function at any generation $g$. However, to avoid numerical overflows caused by multiplication of Boltzmann weights in the matrix elements, it is convenient to rewrite them as

\vspace{-0.5cm}
\begin{equation}
\label{eq7}
f_{g+1}= \frac{4\mathbf{N}_g}{\mathbf{N}_{g+1}}f_g - \frac{2T}{\mathbf{N}_{g+1}}\left[ \ln
\left[1+(Q-1)y_g^2\right] \right ],
\end{equation}

\vspace{-0.5cm}
\begin{equation}
\label{eq7b}
\xi_{g+1}=\xi_{g}\left[ 1 + \frac{\xi_{g}}{\mathbf{D}_{g+1}}\ln
\left[\frac{1+(Q-1)z_g^4}{2+(Q-2)z_g^2} \right]\right]^{-1},\\\\
\end{equation}

\noindent where $f_g=-\frac T{\mathbf{N}_g}\ln a_g$ and $\xi_g=2^g/\ln(\eta_g/\epsilon_g)$ represent, respectively, the free-energy per spin and the correlation length. $\xi_g$ is defined in terms of the $\mathcal{T}_g$ eigenvalues $\eta_g=a_g+(Q-1)b_g$ and $\epsilon_g=a_g-b_g$ ($Q-1$-fold degenerated), while $y_g=b_g/a_g$ and $z_g=(1-y_g)/(1+(Q-1)y_g)$ are auxiliary variables. It is important to recall that, in the $Q\rightarrow1$ limit, Eq. (\ref{eq7a}) leads to a simple relation between $p_0$ and $y_0$ (or $z_0$), namely

\vspace{-0.5cm}
\begin{equation}
\label{eq7c}
p_0=1-y_0=z_0.
\end{equation}

The introduction of extra bonds in the Hamiltonian (\ref{eq8}) does not destroy the scale invariance of the system, so that the TM method can be adapted to include the influence of the new coupling constants $J_g$'s. To this purpose, at each generation $g$, the matrix element $a_g$ must be redefined to account for the new bond that is introduced between the two root sites. Therefore, for $g\geq1$, we have to multiply the matrix element $a_g$ by the Boltzmann weight $\exp(J_g/T)$ so that, for $Q=1$, Eqs. (\ref{eq5a}), (\ref{eq7}), and (\ref{eq7b}) are now written as

\vspace{-0.5cm}
\begin{equation}
\label{eq9a}
a_{g+1} = a_{g}^4 \exp(J_{g+1}/T),
\end{equation}

\begin{equation}
\label{eq9b}
f_{g+1}= \frac{4\mathbf{N}_g}{\mathbf{N}_{g+1}}f_g -
\frac{J_{g+1}}{\mathbf{N}_{g+1}},
\end{equation}

\begin{equation}
\label{eq9c}
\xi_{g+1}=\xi_{g}\left[ 1 + \frac{\xi_{g}}{\mathbf{D}_{g+1}}\ln
\left[\frac{z_g^2}{1 - (1-z_g^2)^2\exp(-J_{g+1}/T)}\right]\right]^{-1}.
\end{equation}

To connect the modified percolation and Hamiltonian models, respectively defined by Eqs. (\ref{eq1b}) and (\ref{eq1c}), and (\ref{eq8}), we require that Eq. (\ref{eq7c}) should be extended to all values of $g\geq1$, namely $p_g=1-y_g=z_g$. If we restrict the analysis to the $A\leq1$ model, this condition is satisfied provided the coupling constants $J_g$ are given by

\begin{equation} \label{eq10}
J_{g+1}=-T\ln \left [1 + \frac{(1-A) (1-y_g)^4}
{(2y_g-y_g^2)^2} \right ].
\end{equation}

\noindent Eq. (\ref{eq10}) warrants that, $\forall g$, the expressions for $p_{g+1}$ and $z_{g+1}$ as function of $p_g$ and $z_g$ are equivalent. It is amazing that the choice of temperature dependent AF coupling constants given by Eq. (\ref{eq10}) leads to a $Q\rightarrow 1$ Potts model that is equivalent to the modified $A\leq1$ percolation model defined by Eqs. (\ref{eq1b}) and (\ref{eq1c}).

Finally, we consider also the magnetization of the Potts model, which is defined as

\vspace{-0.5cm}
\begin{equation} \label{eq11}
m_g = \frac{1+Q\frac{\partial f_g}{\partial h}}{1-Q}.
\end{equation}

\noindent To evaluate $m_g(T,h)$, it is necessary to consider $h\ne 0$ in Eqs. (\ref{eq5}) and (\ref{eq8}). This condition leads to a larger number of distinct matrix elements in the TM, so that the eigenvalues are no longer expressed as simple linear combinations of $a_g$ and $b_g$. In the Appendix we present the complete recurrence maps required for the evaluation of the magnetization. With the help of the identity between $z$ and $p$ stated before, $m_g(T,h=0)$ can be related to the average mass $M_g(p)$ of the percolating cluster when $Q\rightarrow 1$.

\begin{figure}
\includegraphics*[angle=-90,width=8cm]{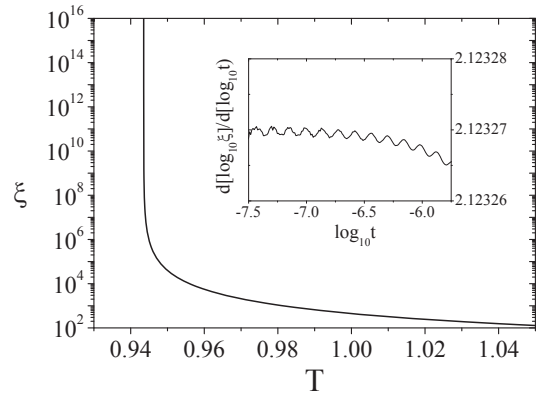}
\caption{Dependence of $\xi$ on $T$ for the $Q\rightarrow1$ modified Potts model ($A=0.9$), with a divergence at $T_c/J_0=0.9435371743..$ . The inset shows the behavior of $d\log_{10}\xi_g/d\log_{10}t$ with respect to $t=|T-T_c|/T_c$, which converges to the exponent $\nu$ when $\log_{10} t\rightarrow -\infty$. Log-periodic oscillations reflect the DHL discrete scale invariance.} \label{fig2}
\end{figure}

The results derived within the TM formalism are in excellent agreement with those obtained with the percolation model also when $A<1$. For the purpose of illustration, we show in Fig. \ref{fig2} the dependence of $\xi_g$ on $T$ for $A=0.9$. We observe a divergence of $\xi$ at $T_c/J_0=0.9435371743...$. For $A=1$, the divergence is observed at $T_c/J_0=1.0390434..=1/\ln(2/(3-\sqrt{5}))$, as predicted by Eq. (\ref{eq7a}). In the inset we draw the dependence of $d\log_{10}\xi_g/d\log_{10}t$, where $t=|T-T_c|/T_c$ is the reduced temperature \cite{Andrade2000}. When $t\rightarrow 0$, this derivative converges to the correct value of $\nu$ already indicated in Fig. \ref{fig4}b. It is remarkable to see that the framework reveals the presence of minute log-periodic oscillations, which are related to the discrete scale invariance of the lattice, even in the $A<1$ cases. Such oscillations are known to be part of the general solution of RG equations, although they can not be evaluated within the linear Jacobian approach.

\begin{figure}
\includegraphics*[angle=-90,width=8cm]{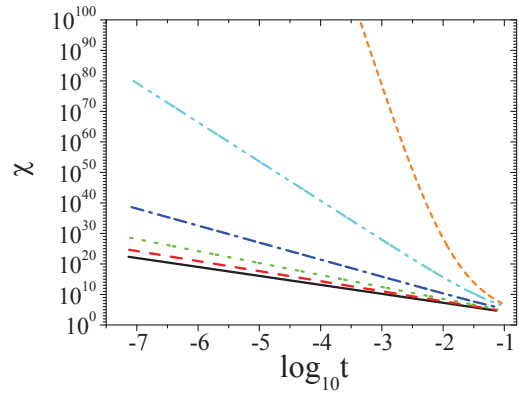}
 \caption{Illustration of the scaling behavior of $\chi$ as function of $t$ (when $T>T_c$) for the following values of $A$: 1.0 (black solid), 0.95 (red dash), 0.9 (green dots), 0.85 (blue dash-dash-dot), 0.82 (cyan dash-dash-dot-dot), 0.815 (orange short dash). The slopes ($\gamma$) increase as $A$ decreases.} \label{fig8}
\end{figure}

The same agreement is noticed in the evaluation of the magnetization $m(T)$. In Fig. \ref{fig60}, where we draw the largest cluster mass $M$ as function of $p$ for the percolation model, we superposed a few points illustrating how $m(T)$ can be transformed into $M(p)$ for $Q\rightarrow1$. To be more precise, Fig. \ref{fig4}c shows the values of the critical exponent $\beta$ obtained from scaling analysis and by the direct slope evaluation of $\log M(p)$ and $\log m(T)$ in the neighborhood of $p_{c,3}$ and $T_c$, respectively.

We finally proceeded with the evaluation of the magnetic susceptibility $\chi_g=d m_g/dh$. As we did not obtain exact expressions for the average cluster size of the modified percolation model, the results we discuss here were evaluated by the numerical iteration of the TM maps for the $Q\rightarrow1$ Potts model only.

For $A=1$, we verified that the critical exponents $\gamma_{-}$ and $\gamma_{+}$, which respectively describe the divergence of $\chi$ for $T<T_c$ and $T>T_c$, coincide within a precision of $10^{-3}$.  However, for $A<1$, the susceptibility behaves differently depending on whether $T<T_c$ or $T>T_c$. In the last case, our results for several values of $A>A_s$ show that $\chi$ obeys a well defined scaling law, which is illustrated by the plots of $\log_{10} \chi \times \log_{10}t$ in Fig.\ref{fig8}. Much as observed with the exponent $\nu$, Fig.\ref{fig8} shows that $\gamma_{+}$ increases when $A$ decreases from 1 to $A_s$.

When $T<T_c$, some subtleties of the model render the numerical evaluation $\gamma_{-}$ almost impossible. First we note that, after a small number of iterations ($g\sim g^{\dag}$), the general feature of $m_g(T<T_c,h=0)$ is to decrease when $g$ increases. This can be understood by the analysis performed in the last Section indicating that, in the neighborhood of $p_{c,4}<1$, $M_{g+1}\simeq o_1 M_g$ with $o_1<1$. In fact, we checked that this equality holds for both the iteration of the maps (\ref{eq4}) and (\ref{eq4a}), as well as for the magnetization recurrence map (\ref{eqA3}) given in the Appendix. The use of the same map to evaluate $m_g(T<T_c,h>0)$ does not show this same behavior. A very small value of $h\sim 10^{-15}$ is sufficient to interrupt the decrease of $m_g$. Therefore, as $g$ increases, the quotient $\Delta m / \Delta h = (m(T<T_c,h>0) - m(T<T_c,h=0))/ h$ increases without bound. Since it was not possible to devise an objective criterion to establish a proper number of interaction steps, we restrict our analyzes to the values obtained for $\gamma_{+}$.

After the independent evaluation of $\nu, \beta,$ and $\gamma_{+}$  we verified whether the equality $d\nu=2\beta + \gamma$ holds when $A\leq1$. The above relation results from a combination of the Rushbrooke and the hyperscale relations, although none of them can be formulated individually. The results shown in Table 1 permit to check whether this equality is verified. Let us remind that the reported values of the exponents $\nu$ and $\beta$ have great accuracy, since they were evaluated by the local properties of the maps (\ref{eq1b}), (\ref{eq1c}), (\ref{eq4}), and (\ref{eq4a}). On the other hand, since the values of $\gamma_{+}$ depend on numerical fittings of the curves shown in Fig.\ref{fig8}, the confidence of the reported values is naturally reduced. The relative error $|\gamma_{-} - 2\nu + 2\beta|/|2\nu - 2\beta|$ increases when $A$ comes close to $A_s$, being however still less than $3\%$.

\begin{table}

\begin{tabular}{|c|c|c|c|c|c|}
\hline
\hline
$A$&$\nu$&$\beta$&$\gamma=2\nu-2\beta$&$\gamma_{+}$&$\Delta\gamma/\gamma$\\  \hline
1    & 1.6353 & 0.1647 & 2.9412  & 2.938   & -1.22$\times10^{-3}$\\
0.95 & 1.8112 & 0.1613 & 3.2996  & 3.293   & -1.99$\times10^{-3}$\\
0.9  & 2.1233 & 0.1567 & 3.9331  & 3.918   & -3.76$\times10^{-3}$\\
0.85 & 2.9602 & 0.1474 & 5.6257  & 5.570   & -9.91$\times10^{-3}$\\
0.82 & 6.6975 & 0.1320 & 13.1311 & 12.842  & -2.20$\times10^{-3}$\\
\hline
\end{tabular}
\caption{Values of the critical exponents of the $A\leq1$ model. The relation $d\nu=2\beta + \gamma$ is found to be satisfied with more than $97.5\%$ accuracy for $A\geq0.82$}\label{table1}
\end{table}

\section{Conclusions}

In this work we considered an alternative path to retard the percolation transition. The approach is based on the use of hierarchical structures satisfying scaling invariance properties, so that renormalization techniques can be applied. The current investigation has been motivated by recent investigations of bond percolation models where the purely random occupation of empty bonds is changed, as to permit a judicious choice of links to be included into the system from a previously selected subset. The strategy used in this work can be better understood in terms of erasing probabilities. Like in the quoted approaches, it retards the emergence of the percolation transition, while the difference between the two strategies is an additional reduction in the number of included bonds. The erasing probabilities described by the parameters $A$, $B$, and $C$ have the effect of changing the universality class of the percolation problems, leading to extreme situations in which the exponent $\nu$ diverges (\emph{singular} transition). For one set of erasing probability we have found that, in spite of the extreme behavior of $\nu$, the non-zero exponent $\beta$ still indicates a continuous transition. For another subset, it is possible to tune the erasing probability in such a way that $\beta$ can be as small as required. This indicates the possibility to have discontinuity in the order parameter. The framework can be further explored to include a richer combination of erasing probabilities, with more than one of the three parameters $A$, $B$, and $C$ being simultaneously smaller than one. We have also shown that an equivalent Potts model with long range interactions can be defined in such a way that, in the $Q\rightarrow1$ limit, it is becomes equivalent to the $A\leq1$ percolation model. This opens another possibility for spin models with discontinuous phase transitions.

{\bf Acknowledgement}: The authors acknowledge support from the European Research Council (ERC) Advanced Grant 319968-FlowCCS, from the Brazilian Agencies FAPESB (project PRONEX 0006/2009) and CNPq, and from the Brazilian National Institute of Science and Technology of Complex Systems (INCT-SC).

\section{Appendix}

The strategy to submit the Potts variables in the DHL to an external field within the TM framework consists in starting with a field free energy at $g=0$. At $g=1$, an external field is introduced to act on the two intermediate sites, but not on the root sites. The same strategy is repeated for each new generation, in such a way that, for any value of $g$, a uniform field acts on all but the root sites. This way, the magnetization can be obtained by deriving the field dependent free energy. It is important to notice that the presence of antiferromagnetic bonds $J_{g}$ with $g\geq1$ causes the response of the Potts variables to a uniform field to become stronger at each generation. Therefore,  it is necessary to reduce the relative magnitude of the applied field at each generation $g$ in order that the value of $M_g(p)$ coincides with $m_g(T,h=0)$.

When $h\neq0$, there exist four different TM elements ($TM_{i,j}$) for any integer value of $Q>2$ at generation $g$: diagonal elements $a_g$ at $i=j=1$ and $c_g$ at $i\geq2, j\geq2$; off-diagonal elements $b_g$ at $i=1,j\geq2$ or $i\geq2,j=1$, and $d_g$ at $i\geq2,j\geq i+1$ or $i\geq3,2\leq j \leq i-1$.

The recurrence relations for this set of matrix elements as a function of $Q$ can be inferred after the explicit evaluation of a few cases of integer values of $Q$. They read

\vspace{-0.5cm}
\begin{equation}
\label{eqA1}
\begin{array}{l}
a_{g+1} = [a_{g}^2v^2+(Q-1)b_{g}^2]^2,
\\
b_{g+1} = b_{g}^2[a_{g}v^2+c_g+(Q-2)d_{g}]^2,
\\
c_{g+1} = [b_{g}^2v^2+c_g^2+(Q-2)d_{g}^2]^2,
\\
d_{g+1} = [b_{g}^2v^2+2c_g d_g + (Q-3)d_{g}^2]^2,
\end{array}
\end{equation}

\noindent where $v^2=\exp(\beta h)$. If we add the extra AF bonds and restrict the analysis to $Q\rightarrow1$ limit, it is possible to show that the recurrence map (\ref{eq9b}) for the free energy $f_g(t,h)=-T\ln(a_g)/N_g$  becomes

\vspace{-0.5cm}
\begin{equation}
\label{eqA2}
f_{g+1}= \frac{4\mathbf{N}_g}{\mathbf{N}_{g+1}}f_g  -
\frac{2T}{\mathbf{N}_{g+1}}\left[ \ln \left[v^2+(Q-1)y_g^2\right] \right ]-
\frac{J_{g+1}}{\mathbf{N}_{g+1}}.
\end{equation}

\noindent After deriving the equation above with respect to $h$, making use of the definition (\ref{eq11}), and taking the limit $Q\rightarrow1$, we obtain

\vspace{-0.5cm}
\begin{equation}
\label{eqA3}
m_{g+1}= \frac{4(\mathbf{N}_g-2)}{\mathbf{N}_{g+1}-2}m_g + \frac{2}{\mathbf{N}_{g+1}-2} + \frac{2y_g
v^{-2}(2T y_g' - y_g)}{\mathbf{N}_{g+1}-2},
\end{equation}

\noindent where $y_g'=dy_g/dh$. The explicit dependence on $\mathbf{N}_g-2$ and $\mathbf{N}_{g+1}-2$ results from the fact the external field does not act on the two root sites. The recurrence relation for $y_{g+1}'=dy_{g+1}/dh$ can be obtained by a straightforward derivation of Eqs.(\ref{eqA1}). It depends on $y_g$ and $y'_g$, as well as on the variables $x_g=c_g/a_g$ and $w_g=d_g/a_g$ and their field derivatives $dx_g/dh$ and $dw_g/dh$. To account for the reduction of the field intensity discussed before, we have to replace the derivative $dy_g/dh=(e^{-J_{g+1}/T})dy_g/dh+y_g d(e^{-J_{g+1}/T})/dh$ by $dy_g/dh=(e^{-J_{g+1}/T})dy_g/dh$. The same procedure should also be used in a similar expression for $dw_g/dh$.

\bibliographystyle{prsty}

\end{document}